\documentclass[aps,prb,twocolumn,showpacs,letterpaper,superscriptaddress,longbibliography]{revtex4-1}

\usepackage[colorlinks=true, citecolor=blue, linkcolor=blue, urlcolor=blue]{hyperref}
\usepackage{graphicx,dcolumn,longtable,epsfig}
\usepackage[usenames]{color}
\usepackage{amssymb}
\usepackage{amsmath}
\usepackage{bm}
\usepackage{footnote}
\usepackage{float}
\usepackage{subfigure}
\usepackage{color}

\usepackage[english]{babel}

\usepackage{ulem}
\usepackage[T1]{fontenc}

\newcommand{\be}{\begin{equation}}
\newcommand{\ee}{\end{equation}}

\def\bea{\begin{eqnarray}}
\def\eea{\end{eqnarray}}

\begin{document}
\title{Effects of phonon dispersion on the bond-bipolaron superconductivity}

\author{Chao Zhang}
\affiliation{Department of Physics, Anhui Normal University, Wuhu, Anhui 241000, China}

\author{Nikolay Prokof'ev}
\affiliation{Department of Physics, University of Massachusetts, Amherst, MA 01003, USA}

\author{Boris Svistunov}
\affiliation{Department of Physics, University of Massachusetts, Amherst, MA 01003, USA}
\affiliation{Wilczek Quantum Center, School of Physics and Astronomy and T. D. Lee Institute, Shanghai Jiao Tong University, Shanghai 200240, China}

\begin{abstract}
We employ the diagrammatic Monte Carlo method 
based on lattice path-integral representaion of the particle sector and real-space diagrammatics 
of the phonon sector to study effects of optical phonon dispersion on Bose-Einstein condensation of bipolarons in the bond model. For dispersionless phonons
with frequency $\omega$ this model was recently shown to give rise to small-size, light-mass bipolarons that undergo a superfluid transition at high values of the $T_c/\omega$ ratio. We find that for dispersive phonons, $T_c$ remains relatively high over a broader range of the coupling strength and even keeps increasing in the deep adiabatic regime both in two and three dimensions. This result implies that phonon dispersion is not a limiting factor in the search for new materials with high superconducting temperatures.
\end{abstract}

\pacs{}
\maketitle

\section{Introduction}
\label{sec:sec1}

Electron-phonon interaction (EPI)  is probably the second most fundamental coupling 
in condensed matter systems and a key mechanism behind superconductivity in metals. 
In the dilute-density regime it leads to formation of polarons
\cite{Landau33,Frohlich50,Feynman:1955du,Schultz:1959el,Holstein59,Alexandrov:1999fy,Holstein2000}, and, ultimately, to bipolarons, which subsequently undergo  a transition to the superconducting state 
(with Bose-Einstein condensation in three dimensions). 
Since binding of polarons takes place in the strong coupling limit, high value of $T_c$ in this scenario requires compact bipolarons with light effective masses. 
This turns out to be impossible to satisfy in the most relevant for materials
adiabatic limit $\omega /t \ll 1$, where $\omega$ is the characteristic phonon frequency 
and $t$ is the electron hopping amplitude, when EPI is of the local density-displacement type 
(Holstein model). The main obstacle in this case is exponentially large
effective bipolaron mass at strong coupling \cite{Chakraverty,PhysRevLett.84.3153,PhysRevB.69.245111}.
This is why recent developments in the field focused on a different EPI mechanism with 
the dominant coupling originating from the phonon-assisted hopping, also known as the bond Su-Schrieffer-Heeger (SSH) interaction. Previous studies of this model considered dispersion-less optical phonons and established that it does feature compact and relatively light  bipolarons \cite{PhysRevB.104.035143,PhysRevB.104.L140307}, opening up the possibility of achieving a high superconducting transition temperature \cite{PhysRevX.13.011010,PhysRevB.108.L220502}.  

Having realistic materials in mind, it is natural to ask to what extent existing results 
apply to dispersive phonons. However, since numerical techniques working in the reduced 
low-energy Hilbert space of the system are not suitable for dealing with delocalized vibration modes, surprisingly few attempts have been made in this direction even for the most thoroughly investigated  Holstein model, see Refs.~\cite{PhysRevB.88.060301, PhysRevB.103.054304}, not to mention that these attempts were limited to one-dimensional chains not representative of realistic materials. 
Moreover, since the adiabatic regime ($\omega/t \ll 1$) is computationally even more 
challenging it has not been investigated in the above mentioned studies. 
Finally, it should be mentioned that the impact of phonon dispersion on the formation of 
charge-density-wave order in the bond SSH model was investigated by the determinant quantum 
Monte Carlo technique \cite{PhysRevLett.120.187003} at high electron density, but this work 
did not explore the low-density limit governed by polarons and bipolarons.
 
In this paper, we employ a newly developed Quantum Monte Carlo (QMC) method based on the path-integral representation of the particle sector and real space diagrammatics of the phonon sector \cite{PhysRevB.105.L020501} to investigate the bipolaron mechanism of superconductivity in the dispersive two- and three-dimensional bond SSH models. To the best of our knowledge, this is the 
first quantitative effort to account for the presence of dispersive phonons in this setup and its effects on the bipolaron binding energies, effective masses, and sizes, as well as estimates of the
highest $T_c$ values within the bipolron mechanism. Overall, we find that the phonon dispersion is not affecting $T_c$ significantly and can even boost $T_c$ to higher values in some parameter regimes. The rest of this paper is organized as follows: In Section \ref{sec:sec2}, we present the Hamiltonian of the bond Su-Schrieffer-Heeger (PSSH) model. In Section \ref{sec:sec3}, we outline the procedure for extracting the bipolaron properties from the Green's function. The results are discussed in Section \ref{sec:sec4}, and conclusions are presented in Section \ref{sec:sec5}.

\section{Model}
\label{sec:sec2}

In this work we focus on the bond SSH model when electronic hopping between the nearest-neighbor (n.n.) sites of the simple cubic/square lattice is modulated by an oscillator located on the bond connecting these sites. The Hamiltonian consists of three terms 
$H=H_{\rm e}+H_{\rm ph}+H_{\rm e-ph}$ representing the free electron, free phonon, and electron-phonon interaction parts, respectively:
\begin{equation}
H_{\rm e} = -t \sum_{\langle i,j \rangle, \sigma}(c_{j, \sigma}^{\dagger}c_{i, \sigma}^{\,} + h. c.) +U \sum_{i} n_{i,\uparrow} n_{j,\downarrow} ,
\label{He}
\end{equation}
\begin{equation}
H_{\rm ph}= \omega_0 \sum_{b} \bigg{(}b^{\dagger}_{b} b_{b }^{\,} +\frac{1}{2} \bigg{)} 
-t_{\rm ph} \sum_{\langle b,b' \rangle} \bigg{(} b^{\dagger}_{b} b_{b' }^{\,} + h.c.    \bigg{)},
\label{Hph}
\end{equation}
\begin{equation}
H_{\rm e-ph}= g  \sum_{b=\langle i, j \rangle, \sigma} \bigg{(} c_{j, \sigma}^{\dagger} c_{i, \sigma}^{\,} + h. c.  \bigg{)} \bigg{(}b_{b}^{\dagger} + b_{b}^{\,} \bigg{)}.
\label{Heph}
\end{equation}
where $b=\langle i, j \rangle$ and $\langle bb' \rangle$ label the n.n. lattice sites (defining the bond $b$) and n.n. bonds oriented in the same direction ($\hat{x}$-, $\hat{y}$- or $\hat{z}$-direction), respectively. Here $c_{i,\sigma}$ ($b_i$) are the electron (optical phonon) annihilation operators on site $i$, $\sigma \in \{ \uparrow, \downarrow \}$ is the spin index, $t$ ($t_{\rm ph}$) is the electron (phonon) hopping amplitude (we take the value of $t$ as the unit of energy), $U$ is the on-site Hubbard repulsion, and $g$ is the hopping-displacement EPI strength. 

The optical phonon spectrum involves two parameters: the bandwidth  $W=4dt_{\rm ph}$, where $d$ is the dimension of space, and the band bottom $\omega_L = \omega_0-W/2$. The value of $t_{\rm ph}$ is chosen to be positive to ensure that the phonon propagator in real space is sign-positive. For direct comparison with the dispersionless case 
we will use the notation $\omega = \omega_L$ if $W>0$ and $\omega=\omega_0$ if $W=0$.  
In terms of the parameter span, we consider the adiabatic ratios $\omega/t$ as small as $0.2375$  with $W/t=0.125$ in 2D
and the phonon bandwidths as large as $W/t=1.2$ for 
$\omega_0/t=1$ in 3D. 

The dimensionless EPI coupling parameter $\lambda$ is defined as
\begin{equation}
\lambda=\frac{g^2}{d t \omega} ,
\label{lambda}
\end{equation}
which is standard in the polaron literature.

\section{Method}
\label{sec:sec3}

Numerically exact solution of model (\ref{He})-(\ref{Heph}) 
is possible by applying the diagrammatic Monte Carlo 
sampling method based on the lattice path-integral representation of the particle sector in combination with the real-space diagrammatic representation of the phonon sector developed in Ref.~\cite{PhysRevB.105.L020501}. The generalization of the 
method to dispersive phonons amounts to simply replacing the local phonon propagator in imaginary time $\tau$,
$D(\mathbf {r}, \tau ) = e^{-\omega \tau} \prod_{i=1}^{d}\delta (r_i)$, with non-local propagator 
$D({\mathbf r}, \tau ) = \prod_{i=1}^{d}I_{r_i} (2t_{\rm ph} \tau)$,
where $I_{n}(z)$ is the modified Bessel function. 

The bipolaron energy $E_{\text{BP}}(\mathbf{k})$ at momentum $\mathbf{k}$ and its typical size can be extracted from the pair Green's function
\begin{equation}
G_{\text{BP}}(\mathbf{R}, \mathbf{r}, \tau ) = 
\langle c_{\mathbf{r}_1, \uparrow}^{\,}(\tau) c_{\mathbf{r}_2, \downarrow}^{\,} (\tau)
c_{0, \downarrow}^{\dagger} (0) c_{0, \uparrow}^{\dagger} (0) \rangle \,.
\label{G2}
\end{equation}
 where $\mathbf{R} =(\mathbf{r}_1+\mathbf{r}_{2})/2 $ and $\mathbf{r} =\mathbf{r}_1-\mathbf{r}_{2} $ are the center of mass and relative distance between the two electrons, respectively.
In the asymptotic  $\tau \to \infty$ limit, the Green's function 
is projected to the ground state in the corresponding momentum sector, as follows from the spectral Lehman representation.
For the stable (non-decaying) quasiparticle state, we have
\begin{equation}
G_{\text{BP}}(\mathbf{k},\mathbf{p}=0, \tau \to \infty) \propto    e^{-[E_{\text{BP}}(\mathbf{k})-\mu] \tau} \, .
\label{Green}
\end{equation}
 
\underline{\textit{Binding energy}}:
The bipolaron binding energy is defined as $\Delta_{\text{BP}}(\mathbf{k})=2 E_{\text{p}}(\mathbf{k})-E_{\text{BP}}(\mathbf{k})$ with $E_{\text{p}}(\mathbf{k})$ the single polaron energy.  

\underline{\textit{Effective mass}}:
The effective mass is extracted from the parabolic dependence 
of the bipolaron energy on momentum at small momenta:  
$E_{\text{BP}}(\mathbf{k}) \approx \mathbf{k}^2/2m_{\text{BP}} + \dots$. 
In practice, low momenta data were fitted to this dependence 
with additional $\propto k^4$ terms to account for higher-order corrections.

\underline{\textit{Mean squared-radius}}:
The advantage of working in the path-integral representaion
for electrons is that ground-state correlators involving particle positions have direct Monte-Carlo estimators. To get the structure of the bound state, we simply collect statistics
for finding two electrons at a distance $\mathbf{r}$ 
in the middle of the large-$\tau $ pair trajectory.
The bipolaron mean squared-radius is then defined as $R^2_{\text{BP}}=\langle (r/2)^2 \rangle =\sum_{\mathbf{r}} (r/2)^2 P(\mathbf{r})$,
where $P(\mathbf{r})$ is the probability of finding two electrons at a given distance in the ground state.   

\underline{\textit{Superfluid transition temperature}}: A dilute gas of bipolarons can be viewed as an interacting bosonic system as long as their density is such that they do not overlap
at the inter-bipolaron distance. At low temperature it undergoes the superfluid transition unless competing instabilities such as phase separation, charge density waves, or Wigner crystallization (in a Coulomb system at extremely low density) 
take place \cite{PhysRevB.104.L201109}. For compact bipolarons, the 
Pauli principle, large Hubbard $U$, and weak Coulomb repulsion 
are all working against the phase separation scenario.

In 2D, bipolarons undergo a superfluid transition at 
$T_ c \approx 1.84 \rho_{\text{BP}}/m_{\text{BP}}$ where $\rho_{\text{BP}}$ is the bipolaron density. This is a robust estimate because $T_c$ dependence on interactions is double-logarithmically weak ~\cite{PhysRevB.37.4936, PhysRevLett.87.270402, PhysRevLett.100.140405}. In 3D, bipolarons undergo the BEC condensation at $T_c \approx 3.2\rho_{\text{BP}}^{2/3}/m_{\text{BP}}$. 
This estimate remains accurate for a remarkably broad range 
of interaction strength (including ideal gas), on a lattice and in continuum, for short and long-range Coulomb potentials
\cite{PhysRevLett.130.236001}. The largest $T_c$ within the bipolaron mechanism thus corresponds to the highest bipolaron
density still satisfying the no-overlap condition.  
This consideration leads the following maximum $T_c$
estimate for the Berezinskii-Kosterlitz-Thouless (BKT) transition in 2D and BEC transition in 3D in the bipolaron liquid~\cite{PhysRevLett.130.236001}
\begin{equation}
T_c \approx 
  \begin{cases}
      & \frac{C}{m_{\text{BP}}^2 R_{\text{BP}}^2}  \quad   \text{if $R_{\text{BP}}^2 \ge 1$} \\
      & \frac{C}{m_{\text{BP}}^2}    \quad \quad \quad  \text{otherwise} ,  \\  
  \end{cases}
  \label{Tc}
\end{equation} 
with $C=0.5$ in 2D and $C=1.2$ in 3D.

\section{Phonon dispersion effects on $T_c$}
\label{sec:sec4}

\begin{figure}[t]
\includegraphics[width=0.45\textwidth]{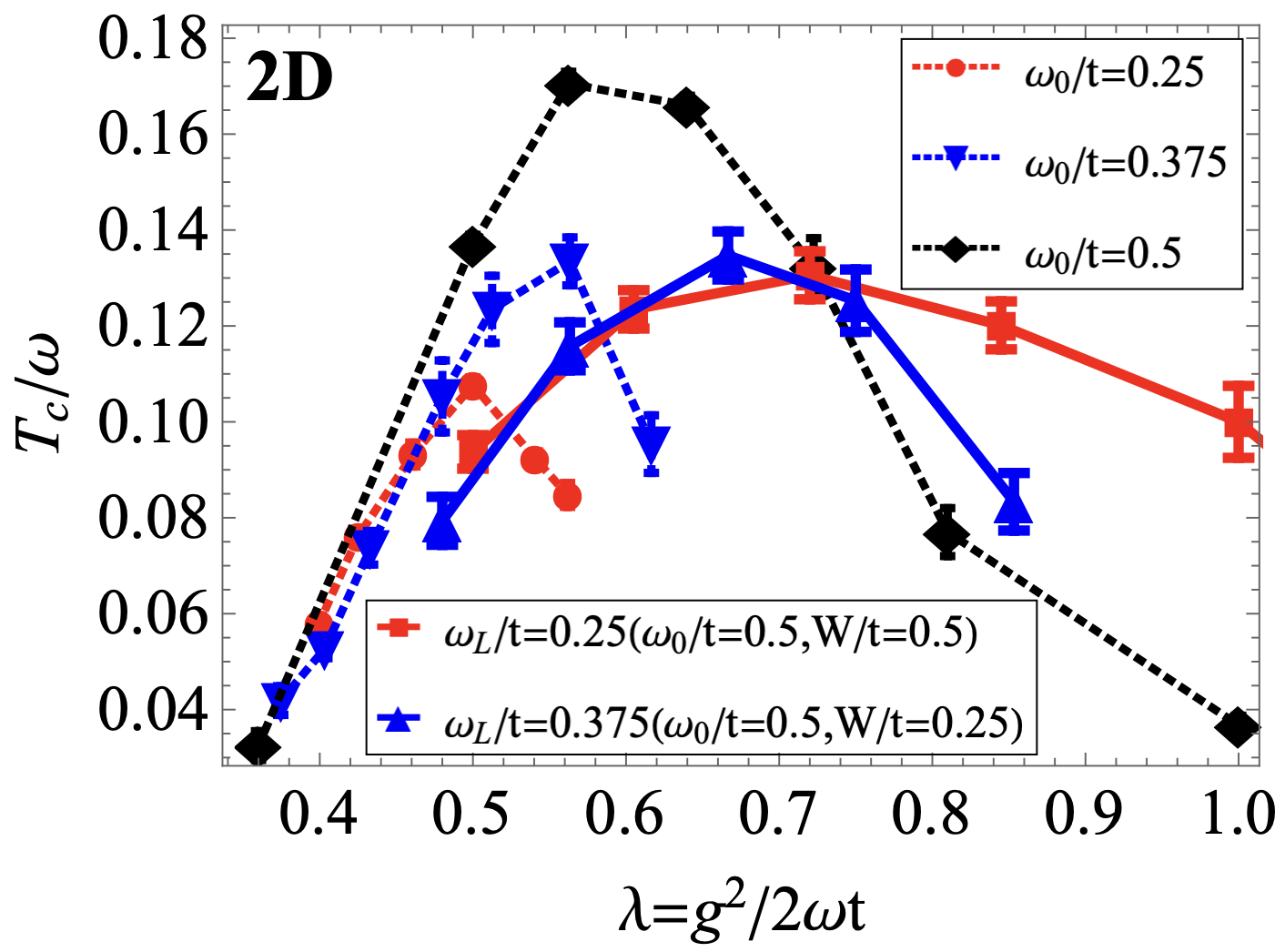}
\caption{Superfluid transition temperature in 2D for adiabatic ratios 
$\omega/t=0.25$ and $\omega/t=0.375$. For the former ratio,
the dispersionless case (red circles connected by a dashed line) is compared with dispersive case for $W/t=0.5$ (red squares connected by a solid line). For the latter ratio, the dispersionless result (blue down-triangles connected by a dashed line) is compared to dispersive case with $W/t=0.25$ (blue up-triangles connected by 
a solid line). By black diamonds connected by a dashed line we show dispersionless results for a larger adiabatic ratio $\omega/t=0.5$, which has the same ``central" phonon frequency as the two dispersive cases described above. The onsite Hubbard repulsion is chosen to be $U=8t$. If not visible, error bars are smaller than symbol size. 
} 
\label{FIG1}
\end{figure}

\begin{figure}[h]
\includegraphics[ width=0.45\textwidth]{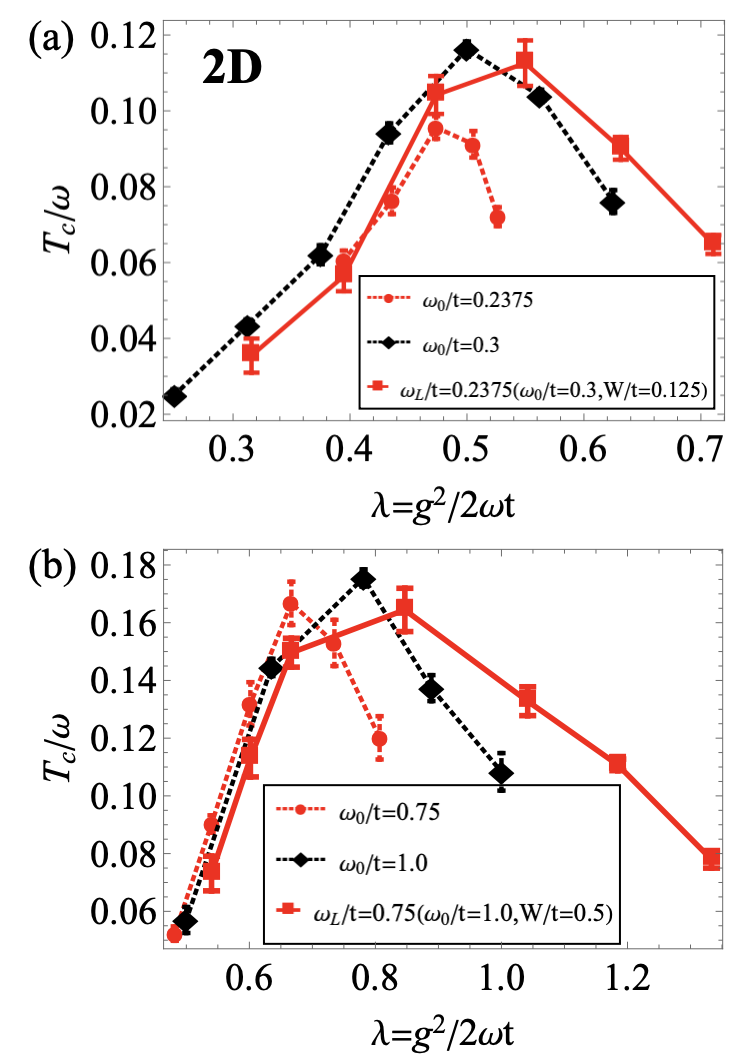}
\caption{(a) Superfluid transition temperature in 2D for adiabatic ratio $\omega/t=0.2375$.
The dispersionless case (red circles connected by a dashed line) is compared with dispersive case for $W/t=0.125$ (red squares connected by a solid line). By black diamonds connected by a dashed line we show dispersionless results for a larger adiabatic ratio $\omega/t=0.3$, which has the same ``central" phonon frequency as the dispersive case described above. 
(b) $T_c/\omega$ ratio in 2D for adiabatic ratio $\omega/t=0.75$.
The dispersionless case (red circles connected by a dashed line) is compared with dispersive case for $W/t=0.5$ (red squares connected by a solid line). By black diamonds connected by a dashed line we show dispersionless results for a larger adiabatic ratio $\omega/t=1.0$, which has the same ``central" phonon frequency as the dispersive case described above. The onsite Hubbard repulsion is chosen to be $U=8t$. If not visible, error bars are smaller than symbol size. 
}
\label{FIG2}
\end{figure}

\begin{figure*}[th]
\includegraphics[width=0.9\textwidth]{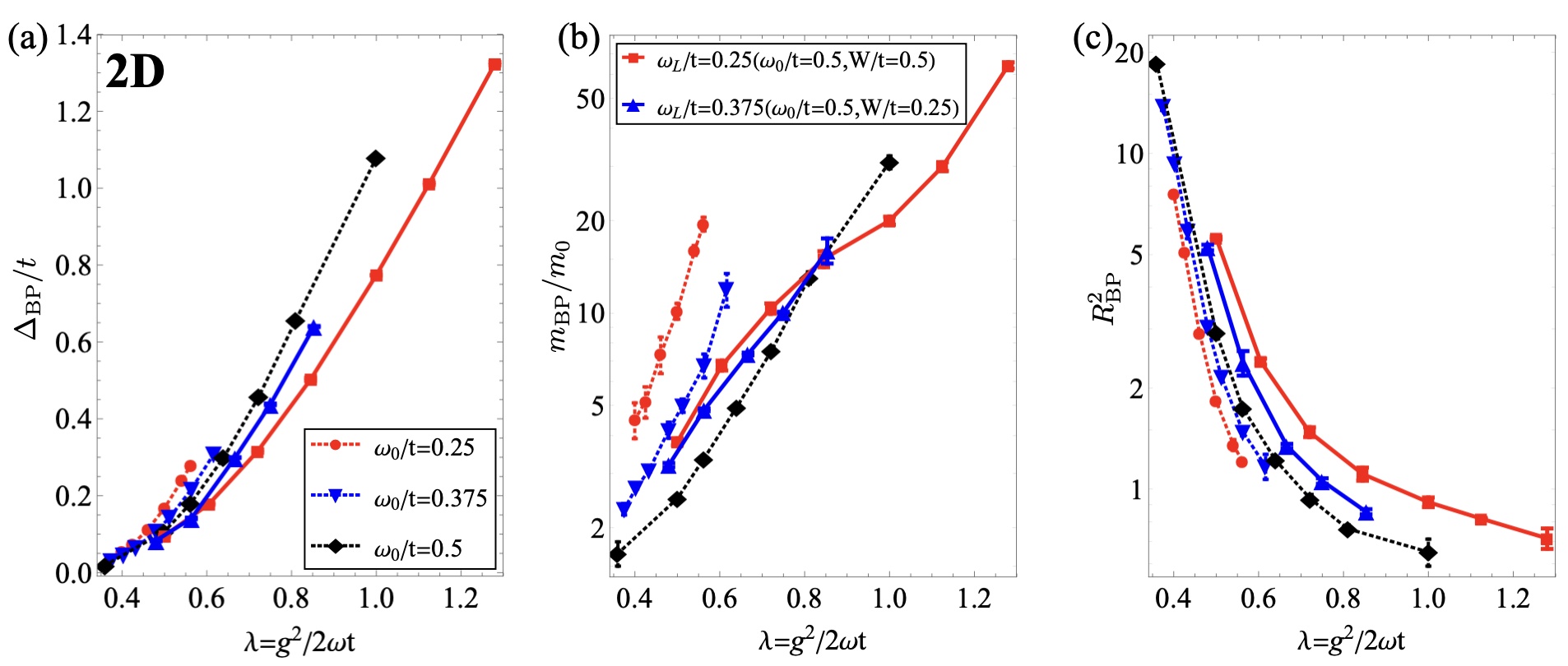}
\caption{Bipolaron properties computed from QMC analysis of Eqs. (\ref{He})-(\ref{Heph}) in 2D 
as functions of $\lambda=g^2/2\omega t$ for adiabatic ratios $\omega/t = 0.25$ and $\omega/t = 0.375$. For the former ratio, the dispersionless case
(red circles connected by a dashed line) is compared with dispersive case for $W/t = 0.5$ (red squares connected by a solid line). For the latter ratio, the dispersionless result (blue down-triangles connected by a dashed line) is compared to dispersive case with $W/t = 0.25$ (blue up-triangles connected by a solid line). By black diamonds connected by a dashed line we show dispersionless results for a larger adiabatic ratio $\omega/t = 0.5$. (a) Bipolaron binding energy $\Delta_{\text{BP}}$ in units of the electron hopping $t$, (b) bipolaron effective mass $m^*_{\text{BP}}$ in units of the mass of two free electrons $m_0=2m_e=1/t$, and (c) bipolaron mean-square radius $R_{\text{BP}}^2$. If not visible, error bars are smaller than symbol size.
}
\label{FIG3}
\end{figure*}

In this section, we consider phonon dispersion effects on the superconducting transition temperature $T_c$ defined by Eq.~(\ref{Tc}) for different adiabatic ratios in both 2D and 3D. The phonon frequencies studied 
here are as low as $\omega=0.2375t$, which is in the deep adiabatic limit relevant for materials. 
Given that the phonon spectrum starts from $\omega_L=\omega_0-2/W$, we compare dispersive results 
to two dispersionless cases with $\omega =\omega_L$ and $\omega =\omega_0$. 
In all cases we choose $U=8t$ to account for strong correlations due to repulsive electron-electron 
interaction. As expected, $T_c/\omega$ has a dome-type shape as a result of competition between the
bipolaron size and its effective mass, see Eq.~(\ref{Tc}), which have opposite dependence on $\lambda$.

\subsection{2D case}

Figures~\ref{FIG1}-\ref{FIG3} present results for $T_c/\omega$ computed from Eq.~(\ref{Tc}) 
(with $m_{\text{BP}}^*$ and $R^2_{\text{BP}}$ parameters extracted from QMC simulations) 
as functions of $\lambda$ for various adiabatic ratios.
In Fig.~~\ref{FIG1} we compare dispersionless models with phonon frequencies 
$\omega_0/t=0.25$ and $\omega_0/t=0.375$, and $\omega_0/t=0.5$ with
models having phonon bandwidths $W/t=0.25t$, and $W/t=0.5t$ and the “central" frequency 
$\omega_0/t=0.5$. 
All three dispersionless cases exhibit a standard non-monotonic, dome-type, dependence of $T_c$
on $\lambda$ \cite{PhysRevX.13.011010,PhysRevB.108.L220502} with the peak value diminishing at 
small $\omega_0$. The same qualitative picture holds for dispersive cases except that the 
region where $T_c/\omega$ is close to a maximum value is more extended and the entire curve 
in this region weakly depends on the $\omega=\omega_L$ value. The peak values are found to be in between the peak values corresponding to dispersionless models with frequencies equal to the 
middle and bottom of the dispersion relation. 

This picture holds from smaller ($\omega/t=0.2375$) and larger ($\omega/t=0.75$) adiabatic ratios, see
Figs.~\ref{FIG2}(a) and \ref{FIG2}(b) where we compare dispersive model results with predictions of two dispersionless models having either the same ``central" frequency or the same lowest frequency. 
We find that  broadening of the region where $T_c$ dependence on the EPI coupling has a maximum
(without loosing the amplitude!) is the most important result of this study because it implies 
that in dispersive models high values of $T_c$ are not limited to a narrow window in the parameter space.
 
In Fig.~\ref{FIG3} we present basic properties of bipolarons which are behind the 
$T_c$ estimates based on Eq.~(\ref{Tc}). Dispersive models tend to have smaller binding energies 
because it is more difficult to gain energy from displacement of harmonic modes with 
large frequencies, see Fig.~\ref{FIG3}(a). As a result, bipolaron states are more extended and 
have larger $R_{\text{BP}}^2$, see Fig.~\ref{FIG3}(c). However, negative effect of larger bipolaron size 
on $T_c$ is compensated by the fact that extended states have smaller effective masses,
see Fig.~\ref{FIG3}(b). 
This effect is very pronounced at strong coupling and small $\omega/t$ because bipolaron motion is 
more difficult to suppress by small oscillator overlap integrals when harmonic modes are delocalized. 
As a result, the region where $T_c/\omega$ curves reach their large values broadens towards larger
$\lambda$; in dispersionless models this parameter space features very large effective masses.

\subsection{3D case} 

\begin{figure}[h]
\includegraphics[ width=0.46\textwidth]{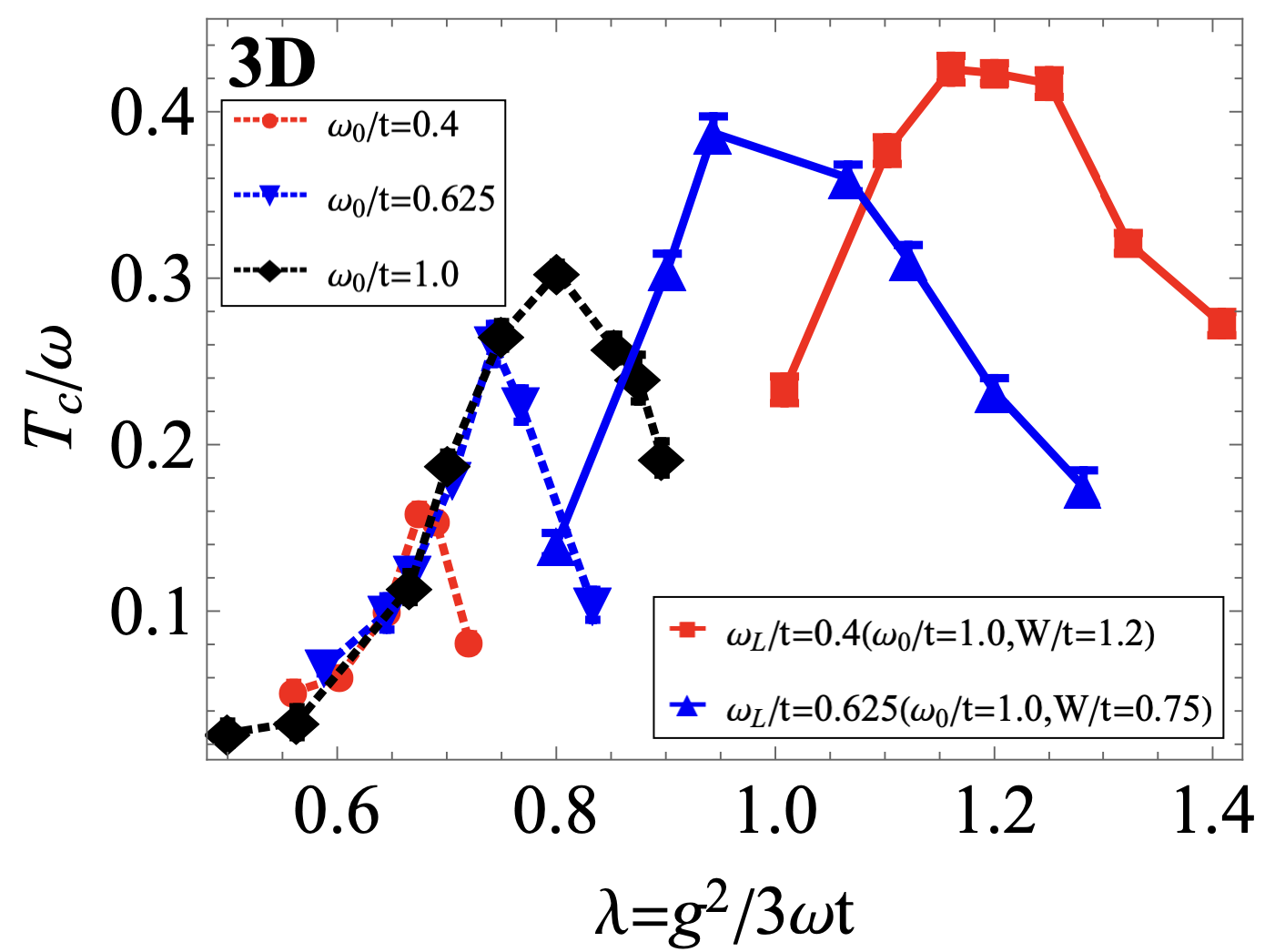}
\caption{Superfluid transition temperature in 3D for adiabatic ratios $\omega/t=0.4$ and $\omega/t=0.625$.
For the former case, the dispersionless curve (red circles connected by a dashed line) is compared with dispersive case for $W/t=1.2$ (red squares connected by a solid line). 
For the latter case, the dispersionless result (blue down-triangles connected by a dashed line) is compared to dispersive case with $W/t=0.75$ (blue up-triangles connected by 
a solid line). By black diamonds connected by a dashed line we show dispersionless results for a larger adiabatic ratio $\omega/t=1.0$, which has the same ``central" phonon frequency as the two dispersive cases described above. The onsite Hubbard repulsion is chosen to be $U=8t$. If not visible, error bars are within symbol size. 
}
\label{FIG4}
\end{figure}

\begin{figure}[th]
\includegraphics[ width=0.46\textwidth]{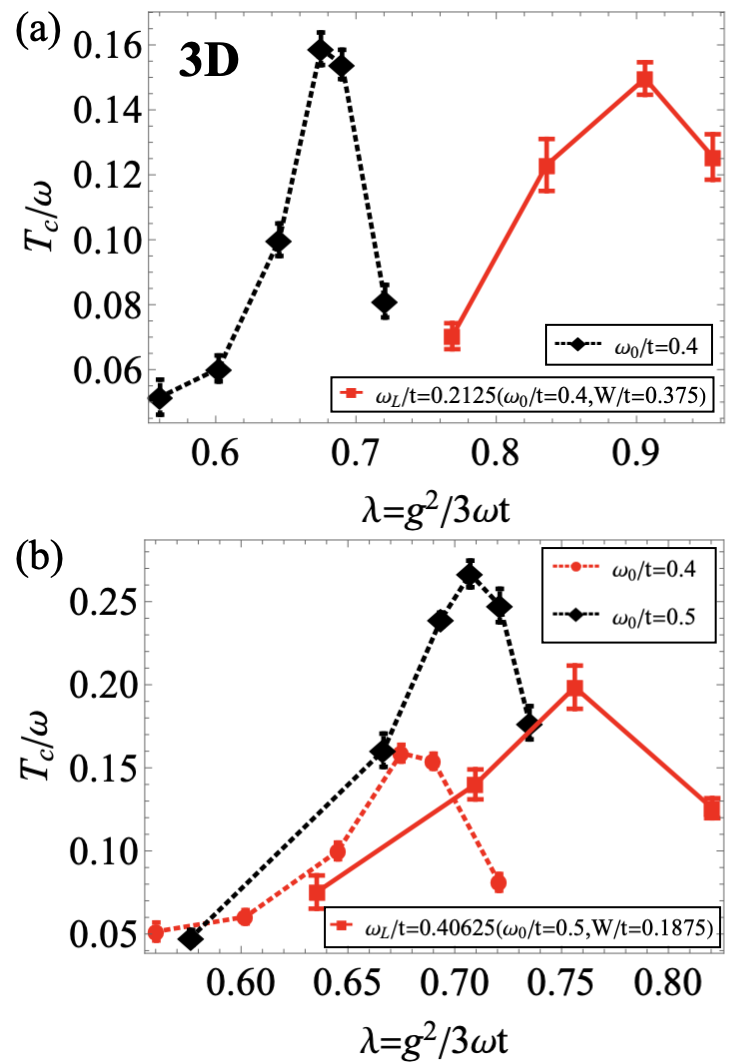}
\caption{(a) Superfluid transition temperature in 3D for adiabatic ratio $\omega/t=0.2125$.
The dispersionless case is not shown here due to computational difficulty in determining exponentially large effective masses. The dispersive case for $W/t=0.375$ (red squares connected by a solid line) is compared to the dispersionless case (black diamonds connected by dashed line) with the same ``central" frequency $\omega/t=0.4$.  
(b) Superfluid transition temperature in 3D for adiabatic ratio $\omega/t=0.40625$.
The dispersionless case (red circles connected by a dashed line) is compared with dispersive case for $W/t=0.1875$ (red squares connected by a solid line). By black diamonds connected by a dashed line we show dispersionless results for a larger adiabatic ratio $\omega/t=0.5$, which has the same ``central" phonon frequency as the dispersive case described above. The onsite Hubbard repulsion is chosen to be $U=8t$. If not visible, error bars are within symbol size. 
}
\label{FIG5}
\end{figure}

Unexpectedly (and in contrast to 2D case) properties of dispersive models in 3D strongly are very sensitive to the adiabatic ratio and the phonon bandwidth. 
For large ``central'' frequency $\omega_0/t=1$, one finds significant 
increase and shift  of $T_c/\omega$ curves towards larger $\lambda$ not only in comparison
with dispersionless models having the same lowest frequency, but also the one with the same central frequency, see Fig.~\ref{FIG4}. 
At small $\omega/t=0.2125$, the dispersive curve with large phonon bandwidth 
has the same amplitude as the dispersionless one with the ``central" frequency
$\omega_0/t=0.4$, but retains a significant shift towards large $\lambda$ and broader character, see Fig.~\ref{FIG5}(a). (When it comes to computing 
properties of dispersionless model with the lowest frequency for this case, the bipolaron effective 
masses are found to be extremely large and hard to determine reliably, implying exponentially small $T_c$ values.) However, if the consider dispersive phonons with smaller phonon bandwidth, then results for 
$T_c$ become similar to what is observed in $2D$: the peak broadens and slightly shifts towards 
larger $\lambda$ while its amplitude is intermediate between the peak values for dispersionless models having the same lowest and ``central" frequencies, see Fig.~\ref{FIG5}(b).  

\begin{figure*}[b]
\includegraphics[width=0.85\textwidth]{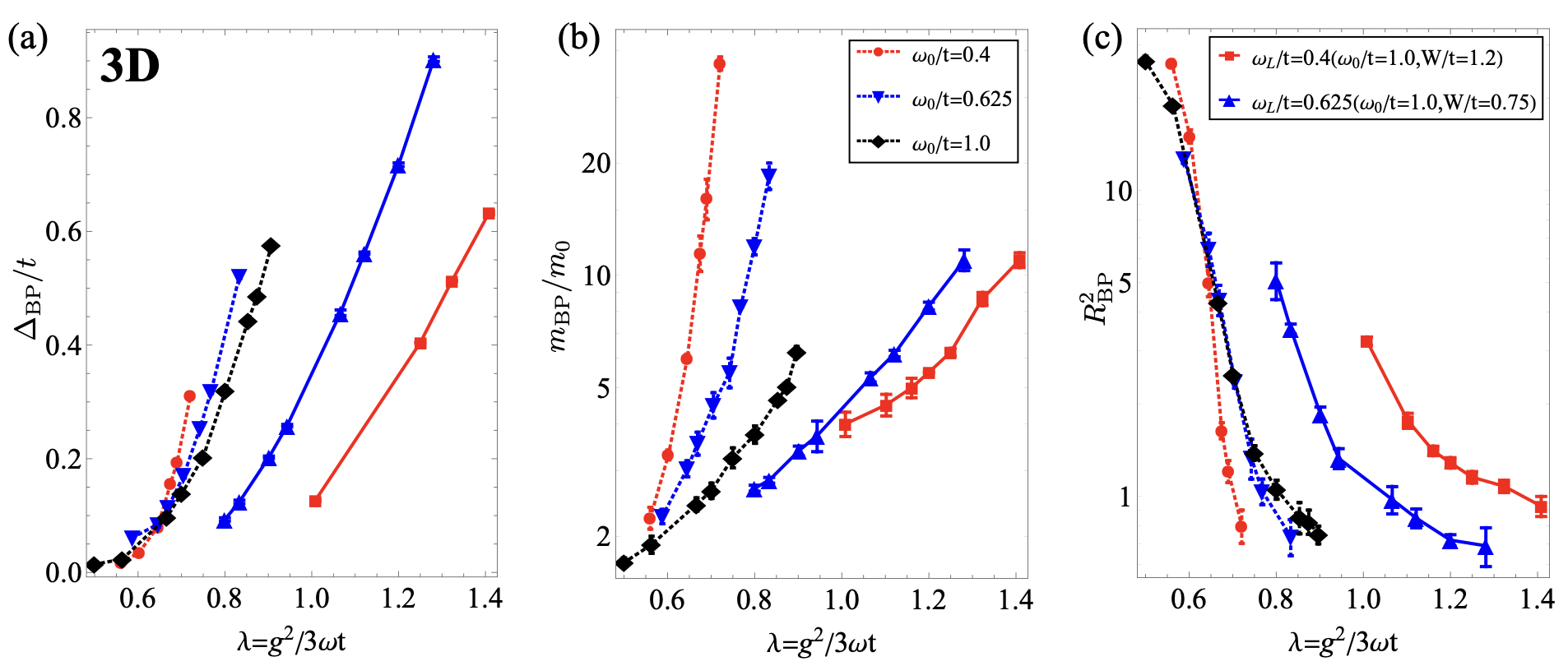}
\caption{Bipolaron properties computed from QMC analysis of Eqs. (\ref{He})-(\ref{Heph}) in 3D 
as functions of $\lambda=g^2/2\omega t$ for adiabatic ratios $\omega/t = 0.4$ and $\omega/t = 0.625$. 
For the former ratio, the dispersionless case (red circles connected by a dashed line) is compared with dispersive case for $W/t = 1.2$ (red squares connected by a solid line).
For the latter ratio, the dispersionless result (blue down-triangles connected by a dashed line) is compared to dispersive case with $W/t = 0.75$ (blue up-triangles connected by a solid line).
By black diamonds connected by a dashed line we show dispersionless results for a larger adiabatic ratio $\omega/t = 1.0$.
(a) Bipolaron binding energy $\Delta_{\text{BP}}$ in units of the electron hopping $t$, (b) bipolaron effective mass $m^*_{\text{BP}}$ in units of the mass of two free electrons $m_0=2m_e=1/t$, and (c) bipolaron mean-square radius $R_{\text{BP}}^2$. If not visible, error bars are smaller than symbol size.
}
\label{FIG6}
\end{figure*}

Figure~\ref{FIG6} compares bipolaron properties (binding energy, effective mass and mean-square radius)
of dispersive and dispersionless models for two adiabatic ratios and phonon bandwidths.
The overall trends (weaker binding energies and more extended states with smaller effective masses for dispersive models) are similar to what was revealed in 2D but in a more dramatic way, as for the same
EPI coupling strength the effective masses and sizes may differ by an order of magnitude. These differences are responsible for strong dependence of final results on the adiabatic ratio and phonon bandwidth even though they are partially cancelled in the $m^*_{\text{BP}}R_{\text{BP}}^2$ product.

\section{Conclusion}
\label{sec:sec5}
Using exact quantum Monte Carlo approach we investigated effects of the optical phonon dispersion on superfluidity of bipolarons in the 2D and 3D bond models. Our results show that under optimal conditions dispersive models still feature small-size light-mass bipolarons with high superfluid transition temperatures. Not only $T_c$ stays relatively high, but also the parameter range 
with near optimal $T_c$ values is getting broader and extends into larger values of the EPI 
coupling strength. This outcomes implies that phonon dispersion is not a limiting factor in the search for new materials with high bipolaron $T_c$.  

\begin{acknowledgments}
BS and NP acknowledge support from the National Science Foundation under Grants DMR-2335904. CZ acknowledges support from the National Natural Science Foundation of China (NSFC) under Grants No 12204173 and 12275002, and the University Annual Scientific Research Plan of Anhui Province under Grant No. 2022AH010013. 
\end{acknowledgments}

\bibliography{dispersive}

\end{document}